\documentclass[11pt]{article}
\usepackage{cite}
\usepackage{mathrsfs}
\usepackage{amssymb,amsfonts,amsmath}
\usepackage{enumerate}
\usepackage{indentfirst}
\usepackage{latexsym,bm}
\usepackage[noend]{algpseudocode}
\usepackage{algorithmicx,algorithm}
\usepackage{algorithm}
\usepackage{makeidx}
\usepackage{fancybox}
\usepackage{color}
\usepackage{multicol}
\usepackage[latin1]{inputenc}
\usepackage{graphicx}
\usepackage{pstricks,pst-node,pst-text,pst-3d}
\usepackage{tikz}
\newtheorem{thm}{Theorem}[section]
\newtheorem{cor}[thm]{Corollary}
\newtheorem{lem}[thm]{Lemma}

\newtheorem{pro}[thm]{Proposition}

\newtheorem{conj}{Conjecture}[section]

\newenvironment{pf}{{\noindent \it \bf Proof:}}{{\hfill$\Box$}\\}

\begin{document}

\title{\bf Strong subgraph $k$-arc-connectivity}
\author{Yuefang Sun$^{1,}$\footnote{Yuefang Sun was supported by National Natural Science Foundation of China (No. 11401389).}{ } and Gregory Gutin$^{2,}$\footnote{Corresponding author. Gregory Gutin was partially supported by Royal Society Wolfson Research Merit Award.} \\
$^{1}$ Department of Mathematics,
Shaoxing University\\
Zhejiang 312000, P. R. China, yuefangsun2013@163.com\\
$^{2}$ Department of Computer Science\\
Royal Holloway, University of London\\
Egham, Surrey, TW20 0EX, UK, g.gutin@rhul.ac.uk}
\date{}
\maketitle

\begin{abstract}
Two previous papers, arXiv:1803.00284 and arXiv:1803.00281, introduced and studied
 strong subgraph $k$-connectivity of digraphs obtaining characterizations, lower and upper bounds and
 computational complexity results for the new digraph parameter. The parameter is an analog of well-studied
 generalized $k$-connectivity of undirected graphs.
 In this paper, we introduce the concept of strong subgraph $k$-arc-connectivity of digraphs, which is an analog of
generalized $k$-edge-connectivity of undirected graphs. We also obtain characterizations, lower and upper bounds and
 computational complexity results for this digraph parameter. Several of our results differ from those obtained for  strong subgraph $k$-connectivity.
\end{abstract}


\section{Introduction}\label{sec:intro}

The generalized $k$-connectivity $\kappa_k(G)$ of a graph $G=(V,E)$
was introduced by Hager \cite{Hager} in 1985 ($2\le k\le |V|$). For
a graph $G=(V,E)$ and a set $S\subseteq V$ of at least two vertices,
an {\em $S$-Steiner tree} or, simply, an {\em $S$-tree}
 is a subgraph
$T$ of $G$ which is a tree with $S\subseteq V(T)$. Two $S$-trees
$T_1$ and $T_2$ are said to be {\em internally disjoint} if
$E(T_1)\cap E(T_2)=\emptyset$ and $V(T_1)\cap V(T_2)=S$. The {\em
generalized local connectivity} $\kappa_S(G)$ is the maximum number
of internally disjoint $S$-trees in $G$. For an integer $k$ with
$2\leq k\leq n$, the {\em generalized $k$-connectivity} is defined
as
$$\kappa_k(G)=\min\{\kappa_S(G)\mid S\subseteq V(G), |S|=k\}.$$
Observe that $\kappa_2(G)=\kappa(G)$. If $G$ is disconnected and
vertices of $S$ are placed in different connectivity components, we
have  $\kappa_S(G)=0$. Thus, $\kappa_k(G)=0$ for a disconnected
graph $G$. Li, Mao and Sun \cite{Li-Mao-Sun} introduced the
following concept of generalized $k$-edge-connectivity. Two
$S$-trees $T_1$ and $T_2$ are said to be {\em edge-disjoint} if
$E(T_1)\cap E(T_2)=\emptyset$ and $V(T_1)\cap V(T_2)\supseteq S$.
The {\em generalized local edge-connectivity} $\lambda_S(G)$ is the
maximum number of edge-disjoint $S$-trees in $G$. For an integer $k$
with $2\leq k\leq n$, the {\em generalized $k$-edge-connectivity} is
defined as
$$\lambda_k(G)=\min\{\lambda_S(G)\mid S\subseteq V(G), |S|=k\}.$$
Observe that $\lambda_2(G)=\lambda(G)$. Generalized connectivity of
graphs has become an established area in graph theory, see a recent
monograph \cite{Li-Mao5} by Li and Mao on generalized connectivity
of undirected graphs.

To extend generalized $k$-connectivity to directed graphs, Sun,
Gutin, Yeo and Zhang \cite{Sun-Gutin-Yeo-Zhang} observed that in the
definition of $\kappa_S(G)$, one can replace ``an $S$-tree'' by ``a
connected subgraph of $G$ containing $S$.'' Therefore,  Sun et al.
\cite{Sun-Gutin-Yeo-Zhang} defined {\em strong subgraph
$k$-connectivity} by replacing ``connected'' with ``strongly
connected'' (or, simply, ``strong'') as follows. Let $D=(V,A)$ be a
digraph of order $n$, $S$ a subset of $V$ of size $k$ and $2\le
k\leq n$. Strong subgraphs $D_1, \dots , D_p$ containing $S$ are
said to be {\em internally disjoint} if $V(D_i)\cap V(D_j)=S$ and
$A(D_i)\cap A(D_j)=\emptyset$ for all $1\le i<j\le p$. Let
$\kappa_S(D)$ be the maximum number of internally disjoint strong
digraphs containing $S$ in $D$. The {\em strong subgraph
$k$-connectivity} is defined as
$$\kappa_k(D)=\min\{\kappa_S(D)\mid S\subseteq V, |S|=k\}.$$
By definition, $\kappa_2(D)=0$ if $D$ is not strong. Sun et al.
\cite{Sun-Gutin-Yeo-Zhang} studied complexity of computing
$\kappa_k(D)$ for arbitrary digraphs, semicomplete
digraphs, and symmetric digraps. In \cite{Sun-Gutin},
Sun and Gutin gave a sharp upper bound for the parameter
$\kappa_k(D)$ and then studied the minimally strong subgraph
$(k,\ell)$-connected digraphs.


As a natural counterpart of the strong subgraph $k$-connectivity, we
now introduce the concept of strong subgraph $k$-arc-connectivity.
Let $D=(V(D),A(D))$ be a digraph of order $n$, $S\subseteq V$ a
$k$-subset of $V(D)$ and $2\le k\leq n$. 
Let $\lambda_S(D)$ be the maximum number of arc-disjoint strong digraphs
containing $S$ in $D$. The {\em strong subgraph
$k$-arc-connectivity} is defined as
$$\lambda_k(D)=\min\{\lambda_S(D)\mid S\subseteq V(D), |S|=k\}.$$
By definition, $\lambda_2(D)=0$ if $D$ is not strong.

A digraph $D=(V(D), A(D))$ is called {\em minimally strong subgraph
$(k,\ell)$-arc-connected} if $\lambda_k(D)\geq \ell$ but for any arc
$e\in A(D)$, $\lambda_k(D-e)\leq \ell-1$.

In this paper, we prove that for fixed integers $k,\ell\ge 2$, the problem of 
deciding whether $\lambda_S(D)\ge \ell$ is NP-complete for a digraph $D$ and a set $S\subseteq V(D)$ of size $k$.
This result is proved in Section \ref{sec:NP} using the corresponding result for $\kappa_S(D)$ proved in \cite{Sun-Gutin-Yeo-Zhang}.
In Section \ref{sec:bounds}, we give lower and upper bounds for the parameter $\lambda_k(D)$ including a lower bound whose analog 
for $\kappa_k(D)$ does not hold as well as Nordhaus-Gaddum type bounds. 

In Section \ref{sec:class} we consider classes of digraphs. 
We characterize when $\lambda_k(D)\ge 2$, $2\le k\le n$, for both semicomplete and symmetric digraphs $D$ of order $n$. The characterizations imply that the problem of deciding whether
$\lambda_k(D)\ge 2$ is polynomial-time solvable for both semicomplete and symmetric digraphs. For fixed $\ell\ge 3$ and $k\ge 2$, the complexity of deciding whether
$\lambda_k(D)\ge \ell$ remains an open problem for both semicomplete and symmetric digraphs. It was proved in   \cite{Sun-Gutin-Yeo-Zhang} that for fixed $k, \ell\ge 2$
the problem of deciding whether $\kappa_k(D)\ge \ell$ is polynomial-time solvable for both semicomplete and symmetric digraphs, but it appears that the approaches 
to prove the two results cannot be used for $\lambda_k(D)$. In fact, we would not be surprised if the $\lambda_k(D)\ge \ell$ problem turns out to be NP-complete 
at least for one of the two classes of digraphs. Also, in Section \ref{sec:class} we prove that $\lambda_2(G\Box H)\geq \lambda_2(G)+ \lambda_2(H),$ where $G\Box H$ is 
the Cartesian product  of digraphs $G$ and $H$. 

Finally, in Section \ref{sec:minimally} we characterize minimally strong subgraph $(2,n-2)$-arc-connected digraphs.
This characterization is different from that of minimally strong subgraph $(2,n-2)$-connected digraphs obtained in \cite{Sun-Gutin}.

\paragraph{Additional Terminology and Notation.} For a digraph $D$, its {\em reverse} $D^{\rm rev}$ is a digraph with same vertex set and such that
$xy\in A(D^{\rm rev})$ if and only if $yx\in A(D)$. A digraph $D$ is
{\em symmetric} if $D^{\rm rev}=D$. In other words, a symmetric digraph $D$ can be obtained from
its underlying undirected graph $G$ by replacing each edge of $G$
with the corresponding arcs of both directions, that is,
$D=\overleftrightarrow{G}.$ A 2-cycle $xyx$ of a strong digraph $D$ is called a {\em bridge} if $D-\{xy,yx\}$ is disconnected. Thus, a bridge corresponds to a bridge in the underlying undirected graph of $D$. An {\em orientation} of a digraph $D$ is a digraph obtained from $D$ by deleting an arc in each 2-cycle of $D$. A digraph $D$ is {\em semicomplete} if for every distinct $x,y\in V(D)$ at least one of the arcs $xy,yx$ in in $D$. Tournaments form a subclass of semicomplete digraphs. A digraph $D$ is $k$-{\em regular} if the in- and out-degree of every vertex of $D$ is equal to $k$. 


\section{NP-completeness}\label{sec:NP}

Yeo proved that it is an NP-complete problem to decide whether a 2-regular digraph has two arc-disjoint hamiltonian
cycles (see, e.g., Theorem 6.6 in \cite{BangY}). Thus, the problem of deciding whether $\lambda_n(D)\ge 2$ is NP-complete, where $n$ is the order of $D$.
We will extend this result in Theorem \ref{thmNP}.

Let $D$ be a digraph and let $s_1,s_2,\ldots{},s_k,t_1,t_2,\ldots{},t_k$ be a collection of not necessarily distinct vertices of $D$.
 A {\em weak $k$-linkage} from $(s_1,s_2,\ldots{},s_k)$ to $(t_1,t_2,\ldots{},t_k)$ is a collection of $k$ arc-disjoint paths
 $P_1,\ldots{},P_k$ such that $P_i$ is
an $(s_i,t_i)$-path for each $i\in [k]$.
A digraph $D=(V,A)$ is
{\em weakly $k$-linked} if it contains a weak $k$-linkage
 from $(s_1,s_2,\ldots{},s_k)$ to $(t_1,t_2,\ldots{},t_k)$ for every choice of (not necessarily
distinct) vertices $s_1,\ldots{},s_k,t_1,\ldots{},t_k$. The {\sc weak
$k$-linkage problem} is the following. Given a digraph $D=(V,A)$
and  distinct vertices $x_1,x_2,\ldots{},x_k,
y_1,y_2,\ldots{},y_k$; decide whether $D$ contains $k$
arc-disjoint paths $P_1,\ldots{},P_k$ such that $P_i$ is an
$(x_i,y_i)$-path. The problem is well-known to be NP-complete already for $k=2$ \cite{Bang-Jensen-Gutin}.

\begin{figure}[tb]
\begin{center}
\tikzstyle{vertexX}=[circle,draw, fill=gray!10, minimum size=12pt, scale=0.8, inner sep=0.3pt]
\begin{tikzpicture}[scale=0.64]
\node (x) at (4.0,4.0) [vertexX] {$x$};
\node (y) at (7.0,4.0) [vertexX] {$y$};
\node (s1) at (1.0,1.0) [vertexX] {$s_1$};
\node (t1) at (4.0,1.0) [vertexX] {$t_1$};
\node (s2) at (7.0,1.0) [vertexX] {$s_2$};
\node (t2) at (10.0,1.0) [vertexX] {$t_2$};
\draw [->, line width=0.03cm] (x) -- (s1);
\draw [->, line width=0.03cm] (t1) -- (x);
\draw [->, line width=0.03cm] (y) -- (s2);
\draw [->, line width=0.03cm] (t2) -- (y);

\draw [->, line width=0.03cm] (x) to [out=330, in=120] (s2);
\draw [->, line width=0.03cm] (s2) to [out=150, in=300] (x);

\draw [->, line width=0.03cm] (y) to [out=210, in=60] (t1);
\draw [->, line width=0.03cm] (t1) to [out=30, in=240] (y);

\draw [rounded corners] (0,-0.5) rectangle (11,2.5);
\node at (12.0,1.0) {$D$};
 \end{tikzpicture}
\end{center}
\caption{The digraph $D'$.} \label{picDp}
\end{figure}
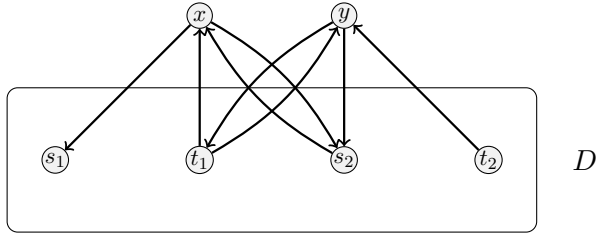

\begin{thm}\label{thmNP}
Let $k\ge 2$ and $\ell\ge 2$ be fixed integers.
Let $D$ be a digraph and $S \subseteq V(D)$ with $|S|=k$. The
problem of deciding whether $\lambda_S(D)\ge \ell$ is NP-complete.
\end{thm}
\begin{pf}
Clearly, the problem is in NP. We will show that it is NP-hard using a reduction similar to that in Theorem 2.1 of  \cite{Sun-Gutin-Yeo-Zhang}.  Let us first deal with the case of $\ell=2$ and $k=2$.
Consider the digraph $D'$ used in the proof of Theorem 2.1 of   \cite{Sun-Gutin-Yeo-Zhang} (see Fig. \ref{picDp}), where $D$ is an arbitrary digraph, $x,y$ are vertices not in $D$, and $t_1x,xs_1, t_2y,ys_2,  xs_2,s_2x,yt_1,t_1y$
are additional arcs. To construct a new digraph $D''$ from $D'$, replace every vertex $u$ of $D$ by two vertices $u^-$ and $u^+$ such that $u^-u^+$ is an arc in $D''$ and for every $uv\in A(D)$ add an arc $u^+v^-$ to $D''$. Also, for $z\in \{x,y\}$, for every arc $zu$ in $D'$ add an arc $zu^-$ to $D''$ and for every arc $uz$ add an arc $u^+z$ to $D''$.

Let $S=\{x,y\}$. It was proved in Theorem 2.1 of
\cite{Sun-Gutin-Yeo-Zhang} that $\kappa_S(D')\ge 2$ if and only if
there are vertex-disjoint paths from $s_1$ to $t_1$ and from $s_2$
to $t_2$. It follows from this result and definition of $D''$ that
$\lambda_S(D'')\ge 2$ if and only if there are
arc-disjoint paths from $s_1^-$ to $t^+_1$ and from $s_2^-$ to
$t^+_2$. Since the {\sc weak 2-linkage problem} is NP-complete, we
conclude that the problem of deciding whether $\lambda_S(D'')\ge 2$
is NP-hard.

Now let us consider the case of $\ell \ge 3$ and $k=2$.
Add to $D''$ $\ell -2$ copies of the 2-cycle $xyx$ and subdivide the arcs of every copy to avoid parallel arcs.
Let us denote the new digraph by $D'''$. Similarly to that in Theorem 2.1 of  \cite{Sun-Gutin-Yeo-Zhang}, we can show that $\lambda_S(D''')\ge \ell$ if and only if $\lambda_S(D'')\ge 2$.

It remains to consider the case of $\ell \ge 2$ and $k\ge 3$. Add to $D'''$ (where $D'''=D''$ for $\ell =2$) $k-2$ new vertices $x_1,\dots ,x_{k-2}$ and arcs of $\ell$ 2-cycles $xx_ix$ for each $i\in [k-2]$.
Subdivide the new arcs to avoid parallel arcs. Let denote the obtained digraph $D''''$. Let $S=\{x,y,x_1,\dots ,x_{k-2}\}$. Similarly to that in Theorem 2.1 of  \cite{Sun-Gutin-Yeo-Zhang}, we can show that $\lambda_S(D'''')\ge \ell$ if and only if $\lambda_S(D'')\ge 2$.
\end{pf}

\section{Bounds for Strong Subgraph $k$-arc-connectivity}\label{sec:bounds}


Let us start this section from observations that can be easily verified using definitions of $\lambda_{k}(D)$ and $\kappa_k(D)$.

\begin{pro}
Let $D$ be a digraph of order $n$, and let $k\ge 2$ be an integer. Then
\begin{equation}\label{monot}
\lambda_{k+1}(D)\leq \lambda_{k}(D) \mbox{ for every } k\le n-1
\end{equation}
\begin{equation}\label{thm1}
\lambda_k(D')\leq \lambda_k(D) \mbox{ where $D'$ is a spanning subgraph of $D$}
\end{equation}
\begin{equation}\label{thm2}
\kappa_k(D)\leq \lambda_k(D) \leq \min\{\delta^+(D), \delta^-(D)\}
\end{equation}
\end{pro}

We will use the following Tillson's decomposition theorem.

\begin{thm}\cite{Tillson}\label{thm01}
The arcs of $\overleftrightarrow{K}_n$ can be decomposed into
Hamiltonian cycles if and only if $n\neq 4,6$.
\end{thm}

Sun et al. obtained the following sharp bounds for
$\kappa_k(D)$.

\begin{thm}\label{thm03}\cite{Sun-Gutin-Yeo-Zhang}
Let $2\leq k\leq n$. For a strong digraph $D$ of order $n$, we have
$$1\leq \kappa_k(D)\leq n-1.$$ Moreover, both bounds are sharp, and
the upper bound holds if and only if $D\cong
\overleftrightarrow{K}_n$, $2\leq k\leq n$ and $k\not\in \{4,6\}$.
\end{thm}

In their proof, they used the following result on $\kappa_k(\overleftrightarrow{K}_n)$.

\begin{lem}\label{thm02} For $2\leq k\leq n$, we have
\[
\kappa_k(\overleftrightarrow{K}_n)=\left\{
   \begin{array}{ll}
      {n-1}, & \mbox{if $k\not\in \{4,6\}$;}\\
     {n- 2}, &\mbox{otherwise.}
   \end{array}
   \right.
\]
\end{lem}

We can now compute the exact values of
$\lambda_k(\overleftrightarrow{K}_n)$.

\begin{lem}\label{thm3} For $2\leq k\leq n$, we have
\[ \lambda_k(\overleftrightarrow{K}_n)=\left\{
   \begin{array}{ll}
      {n-1}, & \mbox{if $k\not\in \{4,6\}$,~or,~$k\in \{4,6\}$~and~$k<n$;}\\
     {n- 2}, &\mbox{if $k=n\in \{4,6\}$.}
   \end{array}
   \right.
\]
\end{lem}

\begin{pf}
For the case that $2\leq k\leq n$~and~$k\not\in \{4,6\}$, by (\ref{thm2}) and
Lemma \ref{thm02}, we have $n-1\leq
\kappa_k(\overleftrightarrow{K}_n)\leq
\lambda_k(\overleftrightarrow{K}_n)\leq n-1$. Hence, $\lambda_k(\overleftrightarrow{K}_n)= n-1$ and
in the following argument we assume that $2\leq k\leq n$~and~$k\in \{4,6\}$.

We first consider the case of $2\leq k=n$. For
$n=4$, since $K_n$ contains a Hamiltonian cycle, the two
orientations of the cycle imply that
$\lambda_n(\overleftrightarrow{K}_n) \geq 2 = n-2$. To see that
there are at most two arc-disjoint strong spanning subgraphs of
$\overleftrightarrow{K}_n$, suppose that there are three
arc-disjoint such subgraphs. Then each such subgraph must have
exactly four arcs (as $|A(\overleftrightarrow{K}_n)|=12$), and so
all of these three subgraphs are Hamiltonian cycles, which means
that the arcs of $\overleftrightarrow{K}_n$ can be decomposed into
Hamiltonian cycles, a contradiction to Theorem~\ref{thm01}). Hence,
$\lambda_n(\overleftrightarrow{K}_n)= n-2$ for $n=4$. Similarly, we
can prove that $\lambda_n(\overleftrightarrow{K}_n)= n-2$ for $n=6$,
as $K_n$ contains two edge-disjoint Hamiltonian cycles, and
therefore $\overleftrightarrow{K}_n$ contains four arc-disjoint
Hamiltonian cycles.

We next consider the case of $2\leq k\leq n-1$. We
assume that $k=6$ as the case of $k=4$ can be considered in a
similar and simpler way. Let $S\subseteq
V(\overleftrightarrow{K}_n)$ be any vertex subset of size six.
Let $S=\{u_i\mid 1\leq i\leq
6\}$ and $V(\overleftrightarrow{K}_n)\setminus S=\{v_j\mid 1\leq
j\leq n-6\}$. Let $D_1$ be the cycle $u_1u_2u_3u_4u_5u_6u_1$; let
$D_2=D_1^{\rm rev}$; let $D_3$ be the cycle
$u_1u_3u_6u_4u_2u_5u_1$; let $D_4=D_3^{\rm rev}$;
let $D_5$ be a subgraph of
$\overleftrightarrow{K}_n$ with vertex set $S\cup \{v_1\}$ and arc
set $\{u_1v_1, v_1u_2, u_2u_6, u_6v_1, v_1u_5, u_5u_3, u_3v_1,
v_1u_4, u_4u_1\}$; let $D_6=D_5^{\rm rev}$;
for each $x\in \{v_j\mid 2\leq j\leq n-6\}$, let
$D_x$ be a subgraph of $\overleftrightarrow{K}_n$ with vertex set
$S\cup \{x\}$ and arc set $\{xu_i, u_ix\mid 1\leq i\leq 6\}$. Hence,
we have $\lambda_S(D)\geq n-1$ for any $S\subseteq
V(\overleftrightarrow{K}_n)$ with $|S|=6$ and so $\lambda_k(D)\geq
n-1$. We clearly have $\lambda_k(D)\leq n-1$ by (\ref{thm2}), then
our result holds.
\end{pf}

Now we  obtain sharp lower and upper bounds for
$\lambda_k(D)$ for $2\leq k\leq n$.

\begin{thm}\label{thma}
Let $2\leq k\leq n$. For a strong digraph $D$ of order $n$, we have
$$1\leq \lambda_k(D)\leq n-1.$$ Moreover, both bounds are sharp, and
the upper bound holds if and only if $D\cong
\overleftrightarrow{K}_n$, where $k\not\in
\{4,6\}$,~or,~$k\in \{4,6\}$~and~$k<n$.
\end{thm}
\begin{pf}
The lower bound is clearly correct by the definition of $\lambda_k(D)$, and
for the sharpness, a cycle is our desired digraph. The upper bound
and its sharpness hold by (\ref{thm1}) and Lemma~\ref{thm3}.

If $D$ is not equal to $\overleftrightarrow{K}_n$ then $\delta^+(D)
\leq n-2$ and by (\ref{thm2}) we observe that $\lambda_k(D) \leq
\delta^+(D) \leq n-2$. Therefore, by Lemma~\ref{thm3}, the upper
bound holds if and only if $D\cong \overleftrightarrow{K}_n$,
where $k\not\in \{4,6\}$,~or,~$k\in
\{4,6\}$~and~$k<n$.
\end{pf}

We can establish the relationship between $\lambda_k(D)$ and
$\lambda(D)$.

\begin{thm}\label{thmb}
For $2\leq k\leq n$, we have
$$\lambda_k(D)\leq \lambda(D).$$ Moreover, the bound is sharp.
\end{thm}
\begin{pf}
Let $A$ be a $\lambda(D)$-arc-cut of $D$, where $1\leq
\lambda(D)\leq n-1$. We choose $S\subseteq V(D)$ such that at least
two of these $k$ vertices are in different strong components of
$D-A$. Thus, any strong subgraph containing $S$ in $D$ must contain
an arc in $A$. By the definition of $\lambda_S(D)$ and
$\lambda_k(D)$, we have $\lambda_k(D)\leq \lambda_S(D)\leq
|A|=\lambda(D)$.

For the sharpness of the bound, 
consider the following digraph $D$ used in the proof
of Theorem 2.2 of \cite{Sun-Gutin}. Let $D$ be a symmetric digraph
whose underlying undirected graph is $K_{k}\bigvee
\overline{K}_{n-k}$~($n\geq 3k$), i.e. the graph obtained from
disjoint graphs $K_{k}$ and $\overline{K}_{n-k}$ by adding all edges
between the vertices in $K_{k}$ and $\overline{K}_{n-k}$. 

Let $V(D)=W\cup U$, where $W=V(K_k)=\{w_i\mid 1\leq i\leq k\}$ and
$U=V(\overline{K}_{n-k})=\{u_j\mid 1\leq j\leq n-k\}$. Let $S$ be any $k$-subset of
vertices of $V(D)$ such that $|S\cap U|=s$ ($s\leq k$) and $|S\cap
W|=k-s$. 
We use the same set of strong subgraphs $D_i$
constructed in the proof of Theorem 2.2 of \cite{Sun-Gutin}. Recall
that $\{D_i\mid 1\leq i\leq k\}$ is a set of $k$ internally disjoint
strong subgraphs containing $S$, so $\lambda_S(D)\geq \kappa_S(D)\geq
k$, and then $\lambda_k(D)\geq k$. Combining this with the bound
that $\lambda_k(D)\leq \lambda(D)$ and the fact that $\lambda(D)\leq
\min\{\delta^+(D), \delta^-(D)\}=k$, we can get $\lambda_k(D)=
\lambda(D)=k$.
\end{pf}

Shiloach \cite{shiloachIPL8} proved the following:

\begin{thm}\cite{shiloachIPL8}
\label{Shiloach}
A digraph $D$ is weakly $k$-linked
 if and only if $D$ is $k$-arc-strong.
\end{thm}

Using Shiloach's Theorem, we will prove the following lower bound for $\lambda_k(D).$ Such a bound does not hold for $\kappa_k(D)$ since it was shown in \cite{Sun-Gutin-Yeo-Zhang} using Thomassen's result in \cite{Thom} that for every $\ell$ there are digraphs $D$ with $\kappa(D)=\ell$ and $\kappa_2(D)=1$.

\begin{pro}\label{lambdas}
Let $k\le \ell=\lambda(D)$. We have $\lambda_k(D)\ge \lfloor \ell/k\rfloor $.
\end{pro}
\begin{pf}
 Choose an arbitrary vertex set $S=\{s_1,\dots ,s_k\}$ of $D$ and let $t=\lfloor \ell/k\rfloor $.  By Theorem \ref{Shiloach}, there is a weak $kt$-linkage $L$ from $x_1,x_2,\ldots{},x_{kt}$ to
$y_1,y_2,\ldots{},y_{kt}$, where $x_i= s_{i \mod k}$ and $y_i=s_{i \mod k +1}$ and $s_{k+1}=s_1$. Note that the paths of $L$ form $t$ arc-disjoint strong subgraphs of $D$ containing $S$.
\end{pf}

For a digraph $D=(V(D), A(D))$, the {\em complement digraph},
denoted by $D^c$, is a digraph with vertex set $V(D^c)=V(D)$ such
that $xy\in A(D^c)$ if and only if $xy\not\in A(D)$.

Given a graph parameter $f(G)$, the Nordhaus-Gaddum Problem is to
determine sharp bounds for (1) $f(G) + f(G^c)$ and (2) $f(G)f(G^c)$,
and characterize the extremal graphs. The Nordhaus-Gaddum type
relations have received wide attention; see a recent survey paper
\cite{Aouchiche-Hansen} by Aouchiche and Hansen. Theorem \ref{thmf}
concerns such type of a problem for the parameter $\lambda_k$. To prove the theorem,
we will need the following.

\begin{pro}\label{thm6}
A digraph $D$ is strong if and only if $\lambda_k(D)\ge 1.$
\end{pro}
\begin{pf} If $D$ is strong, then for
every vertex set $S$ of size $k,$ $D$ has a strong subgraph
containing $S$. If $\lambda_k(D)\ge 1$, for each vertex set $S$ of
size $k$ construct $D_S,$ a strong subgraph of $D$ containing $S.$
The union of all $D_S$ is a strong subgraph of $D$ as there are sets
$S_1, S_2, \dots , S_p$ such that the union of $S_1, S_2, \dots ,
S_p$ is $V(D)$ and for each $i\in [p-1],$ $D_{S_i}$ and
$D_{S_{i+1}}$ share a common vertex.
\end{pf}

\begin{thm}\label{thmf}
For a digraph $D$ with order $n$, the following assertions holds:\\
$(i)$~$0\leq \lambda_k(D)+\lambda_k(D^c)\leq n-1$. Moreover, both bounds are sharp. In particular, the lower bound holds if and only if $\lambda(D)=\lambda(D^c)=0$.\\
$(ii)$~$0\leq \lambda_k(D){\lambda_k(D^c)}\leq (\frac{n-1}{2})^2$.
Moreover, both bounds are sharp. In particular, the lower bound
holds if and only if $\lambda(D)=0$ or $\lambda(D^c)=0$.
\end{thm}
\begin{pf}
We first prove $(i)$. Since $D\cup D^c=\overleftrightarrow{K}_n$, by definition of $\lambda_k$,
$\lambda_k(D)+\lambda_k(D^c)\leq
\lambda_k(\overleftrightarrow{K}_n)$. 
Thus, by Lemma \ref{thm3}, the
upper bound for the sum $\lambda_k(D)+\lambda_k(D^c)$ holds. 
Let $H\cong \overleftrightarrow{K}_n$. When $k\not\in \{4,6\}$,~or,~$k\in
\{4,6\}$~and~$k<n$, by Lemma \ref{thm3}, we have $\lambda_k(H)=n-1$
and we clearly have $\lambda_k(H^c)=0$, so the upper bound is sharp.

The lower bound is clear. Clearly, the lower bound holds, if and only if
$\lambda_k(D)=\lambda_k(D^c)=0$, if and only if
$\lambda(D)=\lambda(D^c)=0$ by Proposition \ref{thm6}.

We now prove $(ii)$. The lower bound is clear. The lower bound
holds, if and only if $\lambda_k(D)=0$ or $\lambda_k(D^c)=0$, if and
only if $\lambda(D)=0$ or $\lambda(D^c)=0$ by Proposition
\ref{thm6}. For the upper bound, we have
$$\lambda_k(D){\lambda_k(D^c)}\leq
\left(\frac{\lambda_k(D)+\lambda_k(D^c)}{2}\right)^2\leq \left(\frac{n-1}{2}\right)^2.$$
Let $H\cong \overleftrightarrow{K}_n$. 
When $k\not\in
\{4,6\}$,~or,~$k\in \{4,6\}$~and~$k<n$, by Lemma \ref{thm3}, we have
$\lambda_k(H)=n-1$ and we clearly have $\lambda_k(H^c)=0$, so the
upper bound is sharp.
\end{pf}


\section{Results for Classes of Digraphs}\label{sec:class}

Bang-Jensen and Yeo \cite{BangY} conjectured the following:

\begin{conj}\label{conj1}
For every $\lambda\ge 2$ there is a finite set ${\cal S}_{\lambda}$ of digraphs such that
$\lambda$-arc-strong semicomplete digraph $D$ contains $\lambda$ arc-disjoint
spanning strong subgraphs unless $D\in {\cal S}_{\lambda}$. 
\end{conj}

Bang-Jensen and Yeo \cite{BangY} proved the conjecture for $\lambda=2$ by showing that $|{\cal S}_2|=1$ and describing the unique digraph $S_4$ of  ${\cal S}_2$ of order 4. This result and Theorem \ref{thmb} imply the following:

\begin{thm}\label{thmT}
For a semicomplete digraph $D$,  of order $n$ and an integer $k$ such that $2\le k\le n$, $\lambda_k(D)\ge 2$ if and only if $D$ is 2-arc-strong and $D\not\cong S_4$.
\end{thm}

Now we turn our attention to symmetric graphs. We start from characterizing symmetric digraphs $D$ with $\lambda_k(D)\ge 2$, an analog of Theorem \ref{thmT}.
 To prove it we will use the following result of Boesch and Tindell \cite{BT} translated from the language of mixed graphs to that of digraphs.

\begin{thm}\label{thm:BT}
A strong digraph $D$ has a strong orientation if and only if $D$ has no bridge.
\end{thm}

Here is our characterization.

\begin{thm}\label{thmSym}
For a strong symmetric digraph $D$ of order $n$ and an integer $k$ such that $2\le k\le n$, $\lambda_k(D)\ge 2$ if and only if $D$ has no bridge.
\end{thm}
\begin{pf}
Let $D$ have no bridge. Then, by Theorem \ref{thm:BT}, $D$ has a strong orientation $H$. Since $D$ is symmetric, $H^{\rm rev}$ is another orientation of $D$. Clearly, $H^{\rm rev}$ is strong and hence $\lambda_k(D)\ge 2$.

Suppose that $D$ has a bridge $xyx$. Choose a set $S$ of size $k$ such that $\{x,y\}\subseteq S$ and observe that any strong subgraph of $D$ containing vertices $x$ and $y$ must include both $xy$ and $yx$. Thus,  $\lambda_S(D)=1$
and $\lambda_k(D)=1$.
\end{pf}

Theorems \ref{thmT} and \ref{thmSym} imply the following complexity result, which we believe to be extendable from $\ell=2$ to any natural $\ell$. 

\begin{cor}\label{cor:poly}
The problem of deciding whether $\lambda_k(D)\ge 2$ is polynomial-time solvable if $D$ is either semicomplete or symmetric digraph of order $n$ and $2\le k\le n. $
\end{cor}

Now we give a lower bound on $\lambda_k(D)$ for symmetric digraphs $D$. 

\begin{thm}\label{thmc}
For every graph $G$, we have
$$\lambda_k(\overleftrightarrow{G})\geq \lambda_k(G).$$ Moreover, this
bound is sharp. In particular, we have
$\lambda_2(\overleftrightarrow{G})=\lambda_2(G)$.
\end{thm}
\begin{pf}
We may assume that $G$ is a connected graph. Let $D$ be a digraph
whose  underlying undirected graph is $G$ and let $S=\{x,y\}$, where
$x,y$ are distinct vertices of $D$. Observe that $\lambda_S(G)\ge
\lambda_S(D)$. Indeed, let $p=\lambda_S(D)$ and let $D_1,\dots ,D_p$
be $S$-arc-disjoint strong subgraphs of $D$. Thus,  by choosing a
path from $x$ to $y$ in each $D_i$, we obtain $p$ arc-disjoint paths
from $x$ to $y$, which correspond to $p$ arc-disjoint paths between
$x$ and $y$ in $G$. Thus, $\lambda (G)=\lambda_2(G)\ge \lambda_2(D)$. 

We now consider the general $k$. Let
$\lambda_S(\overleftrightarrow{G})=\lambda_k(\overleftrightarrow{G})$
for some $S\subseteq V(\overleftrightarrow{G})$ with $|S|=k$. We
know that there are at least $\lambda_k(G)$ edge-disjoint trees
containing $S$ in $G$, say $T_i~(i\in [\lambda_k(G)])$. For each
$i\in [\lambda_k(G)]$, we can obtain a strong subgraph containing
$S$, say $D_i$, in $\overleftrightarrow{G}$ by replacing each edge
of $T_i$ with the corresponding arcs of both directions. Clearly,
any two such subgraphs are arc-disjoint, so we have
$\lambda_k(\overleftrightarrow{G})=\lambda_S(\overleftrightarrow{G})\geq
\lambda_k(G)$, and we also have
$\lambda_2(\overleftrightarrow{G})=\lambda_2(G)=\lambda (G)$.

For the sharpness of the bound, consider the tree $T$ with order
$n$. Clearly, we have $\lambda_k(T)=1$. Furthermore, $1\leq
\lambda_k(\overleftrightarrow{T})\leq \min\{\delta^+(D),
\delta^-(D)\}=1$ by (\ref{thm2}).
\end{pf}

Note that for the case that $3\leq k\leq n$, the equality
$\lambda_k(\overleftrightarrow{G})=\lambda_k(G)$ does not always
hold. For example, consider the cycle $C_n$ of order $n$; it is not
hard to check that $\lambda_k(\overleftrightarrow{C}_n)=2$, but
$\lambda_k(C_n)=1$.

Theorem~\ref{thmc} immediately implies the next result, which follows
from the fact that $\lambda(G)$ can be computed in polynomial time.

\begin{cor}\label{thm4}
For a symmetric digraph $D$, $\lambda_2(D)$ can be computed
in polynomial time.
\end{cor}

Corollaries \ref{cor:poly} and \ref{thm4} shed some light on the complexity of deciding, for fixed $k,\ell\ge 2$, whether $\lambda_k(D)\ge \ell$ for semicomplete and symmetric digraphs $D$. 
However, it is unclear what is the complexity above for every fixed $k,\ell\ge 2$. If Conjecture \ref{conj1} is correct, then the  $\lambda_k(D)\ge \ell$ problem can be solved in polynomial time
for semicomplete digraphs. However, Conjecture \ref{conj1} seems to be very difficult. It was proved in   \cite{Sun-Gutin-Yeo-Zhang} that for fixed $k, \ell\ge 2$
the problem of deciding whether $\kappa_k(D)\ge \ell$ is polynomial-time solvable for both semicomplete and symmetric digraphs, but it appears that the approaches 
to prove the two results cannot be used for $\lambda_k(D)$. Some well-known results such as the fact that the hamiltonicity problem is NP-complete for undirected 3-regular graphs, 
indicate that  the  $\lambda_k(D)\ge \ell$ problem for symmetric digraphs may be NP-complete, too.



In the remainder of this section, we will discuss Cartesian products of digraphs. The
{\em Cartesian product} $G\Box H$ of two digraphs $G$ and $H$ is a
digraph with vertex set
$$V(G\Box H)=V(G)\times V(H)=\{(x, x')\mid x\in V(G), x'\in V(H)\}$$
and arc set $$A(G\Box H)=\{(x,x')(y,y')\mid xy\in A(G),
x'=y',~or~x=y, x'y'\in A(H)\}.$$ By definition, we know the
Cartesian product is associative and commutative, and $G\Box H$ is
strongly connected if and only if both $G$ and $H$ are strongly
connected \cite{Hammack}.

\begin{figure}[!hbpt]
\begin{center}
\includegraphics[scale=0.8]{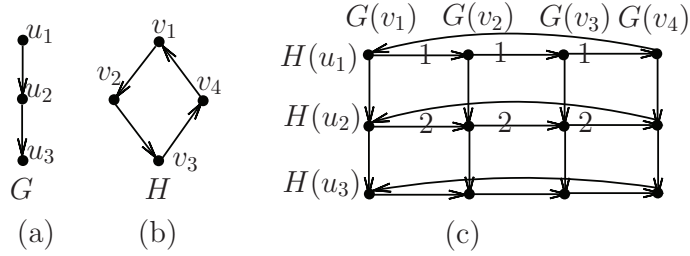}
\end{center}
\caption{Two digraphs $G$, $H$ and their Cartesian
product.}\label{figure1}
\end{figure}

\begin{thm}\label{thmd}
Let $G$ and $H$ be two digraphs. We have $$\lambda_2(G\Box H)\geq
\lambda_2(G)+ \lambda_2(H).$$ Moreover, the bound is sharp.
\end{thm}
\begin{pf}
Let $G$ and $H$ be two digraphs with $V(G)=\{u_i \mid 1\leq i\leq
n\}$ and $V(H)=\{v_j \mid 1\leq j\leq m\}$. We use $G(v_j)$ to
denote the subdigraph of $G\Box H$ induced by vertex set
$\{(u_i,v_j)\mid 1\leq i\leq n\}$ where $1\leq j\leq m$, and use
$H(u_i)$ to denote the subdigraph of $G\Box H$ induced by vertex set
$\{(u_i,v_j)\mid 1\leq j\leq m\}$ where $1\leq i\leq n$. Clearly, we
have $G(v_j)\cong G$ and $H(u_i)\cong H$. (For example, as shown in
Figure \ref{figure1}, $G(v_j)\cong G$ for $1\leq j\leq 4$ and
$H(u_i)\cong H$ for $1\leq i\leq 3$.) For $1\leq j_1\neq j_2\leq m$,
the vertices $(u_i,v_{j_1})$ and $(u_i,v_{j_2})$ belong to the same
digraph $H(u_i)$ where $u_i\in V(G)$; we call $(u_i,v_{j_2})$ the
{\em vertex corresponding to} $(u_i,v_{j_1})$ in $G(v_{j_2})$; for
$1\leq i_1\neq i_2\leq n$, we call $(u_{i_2},v_j)$ the vertex
corresponding to $(u_{i_1},v_j)$ in $H(u_{i_2})$. Similarly, we can
define the subgraph {\em corresponding} to some subgraph. For example,
in the digraph (c) Fig. \ref{figure1}, let $P_1$~$(P_2)$ be the
path labelled 1 (2) in $H(u_1)~(H(u_2))$, then $P_2$ is called the
path {\em corresponding} to $P_1$ in $H(u_2)$.

It suffices to show that there are at least $\lambda_2(G)+
\lambda_2(H)$ arc-disjoint strong subgraphs containing $S$ for any
$S\subseteq V(G\Box H)$ with $|S|=2$. Let $S=\{x, y\}$ and 
consider the following two cases.

{\em Case 1.} $x$ and $y$ are in the same $H(u_i)$ or $G(v_j)$ for
some $1\leq i\leq n, 1\leq j\leq m$. Without loss of generality, we may
assume that $x=(u_1,v_1), y=(u_1,v_2)$. We know there are at least
$\lambda_2(H)$ arc-disjoint strong subgraphs containing $S$ in the
subgraph $H(u_1)$, and so it suffices to find the other $\lambda_2(G)$
strong subgraphs containing $S$ in $G\Box H$.

We know there are at least $\lambda_2(G)$ arc-disjoint strong
subgraphs, say $D_i(v_1)~(i\in [\lambda_2(G)])$, containing the
vertex set $\{x, (u_2,v_1)\}$ in $G(v_1)$. For each $i\in
[\lambda_2(G)]$, we can choose an out-neighbor, say
$(u_{t_i},v_1)$~$(i\in [\lambda_2(G)])$, of $x$ in $D_i(v_1)$ such
that these out-neighbors are distinct. Then in $H(u_{t_i})$, we know
there are $\lambda_2(H)$ arc-disjoint strong subgraphs containing
the vertex set $\{(u_{t_i}, v_1), (u_{t_i}, v_2)\}$, we choose one
such strong subgraph, say $D(H(u_{t_i}))$. For each $i\in
[\lambda_2(G)]$, let $D_i(v_2)$ be the strong subgraph (containing
the vertex set $\{(u_{t_i}, v_2), y\}$) corresponding to $D_i(v_1)$
in $G(v_2)$. We now construct the remaining $\lambda_2(G)$ strong
subgraphs containing $S$ by letting $D_i=D_i(v_1)\cup
D(H(u_{t_i}))\cup D_i(v_2)$ for each $i\in [\lambda_2(G)]$.
Combining the former $\lambda_2(H)$ arc-disjoint strong subgraphs
containing $S$, we can get $\lambda_2(G)+ \lambda_2(H)$ strong
subgraphs, and it is not hard to check that these strong subgraphs
are arc-disjoint.

{\em Case 2.} $x$ and $y$ belong to distinct $H(u_i)$ and $G(v_j)$.
Without loss of generality, we may assume that $x=(u_1,v_1),
y=(u_2,v_2)$.

There are at least $\lambda_2(G)$ arc-disjoint strong subgraphs, say
$D_i(v_1)~(i\in [\lambda_2(G)])$, containing the vertex set $\{x,
(u_2,v_1)\}$ in $G(v_1)$. For each $i\in [\lambda_2(G)]$, we can
choose an out-neighbor, say $(u_{t_i},v_1)$~$(i\in [\lambda_2(G)])$,
of $x$ in $D_i(v_1)$ such that these out-neighbors are distinct.
Then in $H(u_{t_i})$, we know that there are $\lambda_2(H)$ arc-disjoint
strong subgraphs containing the vertex set $\{(u_{t_i}, v_1),
(u_{t_i}, v_2)\}$; we choose one such strong subgraph, say
$D(H(u_{t_i}))$. For each $i\in [\lambda_2(G)]$, let $D_i(v_2)$ be
the strong subgraph (containing the vertex set $\{(u_{t_i}, v_2),
y\}$) corresponding to $D_i(v_1)$ in $G(v_2)$. We now construct the
$\lambda_2(G)$ strong subgraphs containing $S$ by letting
$D_i=D_i(v_1)\cup D(H(u_{t_i}))\cup D_i(v_2)$ for each $i\in
[\lambda_2(G)]$.

Similarly, there are at least $\lambda_2(H)$ arc-disjoint strong
subgraphs, say $D'_j(u_1)~(j\in [\lambda_2(H)])$, containing the
vertex set $\{x, (u_1,v_2)\}$ in $H(u_1)$. For each $j\in
[\lambda_2(H)]$, we can choose an out-neighbor, say
$(u_1,v_{t'_j})$~$(j\in [\lambda_2(H)])$, of $x$ in $D'_j(u_1)$ such
that these out-neighbors are distinct. Then in $G(v_{t'_j})$, we
know there are $\lambda_2(G)$ arc-disjoint strong subgraphs
containing the vertex set $\{(u_1, v_{t'_j}), (u_2, v_{t'_j})\}$, we
choose one such strong subgraph, say $D(G(v_{t'_j}))$. For each
$j\in [\lambda_2(H)]$, let $D'_j(u_2)$ be the strong subgraph
(containing the vertex set $\{(u_2, v_{t'_j}), y\}$) corresponding
to $D'_j(u_1)$ in $H(u_2)$. We now construct the other
$\lambda_2(H)$ strong subgraphs containing $S$ by letting
$D'_j=D'_j(u_1)\cup D(G(v_{t'_j}))\cup D'_j(u_2)$ for each $j\in
[\lambda_2(H)]$.

{\em Subcase 2.1.} $t_i\neq 2$ for any $i\in [\lambda_2(G)]$ and
$t'_j\neq 2$ for any $j\in [\lambda_2(H)]$, that is, $(u_2,v_1)$ was
not chosen as an out-neighbor of $(u_1,v_1)$ in $G(v_1)$ and
$(u_1,v_2)$ was not chosen as an out-neighbor of $(u_1,v_1)$ in
$H(u_1)$. We can check the above $\lambda_2(G)+ \lambda_2(H)$ strong
subgraphs are arc-disjoint.

{\em Subcase 2.2.} $t_i=2$ for some $i\in [\lambda_2(G)]$ or
$t'_j=2$ for some $j\in [\lambda_2(H)]$, that is, $(u_2,v_1)$ was
chosen as an out-neighbor of $(u_1,v_1)$ in $G(v_1)$ or $(u_1,v_2)$
was chosen as an out-neighbor of $(u_1,v_1)$ in $H(u_1)$. Without
loss of generality, we may assume that $t_1=2$ and $t'_1=2$. We replace
$D_1$, $D'_1$ by $\overline{D}_1$, $\overline{D'}_1$, respectively
as follows: let $\overline{D}_1= D_1(v_1)\cup D(H(u_{t_1}))$ and
$\overline{D}_2= D'_1(u_1)\cup D_1(v_2)$. We can check that the
current $\lambda_2(G)+ \lambda_2(H)$ strong subgraphs are
arc-disjoint.

Hence, the bound holds. For the sharpness of the bound, consider the
Cartesian product $D$ of two dicycles $\overrightarrow{C}_n$ and
$\overrightarrow{C}_m$. By (\ref{thm2}) and the bound, we have
$2=\min\{\delta^+(D), \delta^-(D)\}\geq
\lambda_2(\overrightarrow{C}_n \Box \overrightarrow{C}_m)\geq
\lambda_2(\overrightarrow{C}_n)+\lambda_2(\overrightarrow{C}_m)=2$.
This completes the proof.
\end{pf}

\begin{figure}[htbp]
{\tiny
\begin{center}
\renewcommand\arraystretch{3.5}
\begin{tabular}{|p{1.5cm}|p{1.5cm}|p{1.5cm}|p{1.5cm}|p{1.5cm}|}
\hline & $\overrightarrow{C}_m$ & $\overleftrightarrow{C}_m$ &
$\overleftrightarrow{T}_m$ & $\overleftrightarrow{K}_m$
\\\hline

$\overrightarrow{C}_n$ & $2$ & $3$ & $2$ & $m$
\\\hline

$\overleftrightarrow{C}_n$ & $3$ & $4$ & $3$ & $m+1$
\\\hline

$\overleftrightarrow{T}_n$ & $2$ & $3$ & $2$ & $m$
\\\hline

$\overleftrightarrow{K}_n$ & $n$ & $n+1$ & $n$ & $n+m-2$
\\\hline

\end{tabular}
\vspace*{40pt}

\centerline{\normalsize Table $1$. Precise values for the strong
subgraph 2-arc-connectivity of some special cases.}
\end{center}}
\end{figure}

By (\ref{thm2}) and Theorem \ref{thmd}, we can obtain precise
values for the strong subgraph 2-arc-connectivity of the Cartesian
product of some special digraphs, as shown in the Table. Note that
$\overleftrightarrow{T}_m$ is the symmetric digraph whose underlying undirected graph is a tree
of order $m$.


\section{Minimally Strong Subgraph $(k,\ell)$-arc-connected Digraphs}\label{sec:minimally}

A digraph $D$ is {\em minimally strong} if $D$ is strong but $D-e$
is not for every arc $e$ of $D$. By Proposition \ref{thm6} and
Theorem \ref{thma}, we have the following result.

\begin{pro}\label{thm5}
The following assertions hold:\\
$(i)$~A digraph $D$ is minimally strong subgraph
$(k,1)$-arc-connected
if and only if $D$ is minimally strong digraph;\\
$(ii)$~Let $2\leq k\leq n$. If $k\not\in
\{4,6\}$,~or,~$k\in \{4,6\}$~and~$k<n$, then a digraph $D$ is
minimally strong subgraph $(k,n-1)$-arc-connected if and only if
$D\cong \overleftrightarrow{K}_n$.
\end{pro}

The following result characterizes minimally strong subgraph
$(2,n-2)$-arc-connected digraphs. This characterization is different from the characterization of
minimally strong subgraph
$(2,n-2)$-connected digraphs obtained in \cite{Sun-Gutin}.

\begin{thm}\label{thme}
A digraph $D$ is minimally strong subgraph $(2,n-2)$-arc-connected
if and only if $D$ is a digraph obtained from the complete digraph
$\overleftrightarrow{K}_n$ by deleting an arc set $M$ such that
$\overleftrightarrow{K}_n[M]$ is a union of vertex-disjoint cycles
which cover all but at most one vertex of
$\overleftrightarrow{K}_n$.
\end{thm}

\begin{pf}
Let $D$ be a digraph obtained from the complete digraph
$\overleftrightarrow{K}_n$ by deleting an arc set $M$ such that
$\overleftrightarrow{K}_n[M]$ is a union of vertex-disjoint cycles
which cover all but at most one vertex of
$\overleftrightarrow{K}_n$. To prove the theorem it suffices to show that
(a) $D$ is minimally strong subgraph
$(2,n-2)$-arc-connected, that is, $\lambda_2(D)\geq n-2$ but for any
arc $e\in A(D)$, $\lambda_2(D-e)\leq n-3$, and (b) if a digraph $H$ minimally strong subgraph
$(2,n-2)$-arc-connected then it must be constructed from
$\overleftrightarrow{K}_n$ as the digraph $D$ above.
Thus, the remainder of the proof has two parts.

\paragraph{Part (a).} We just consider the case that
$\overleftrightarrow{K}_n[M]$ is a union of vertex-disjoint cycles
which cover all vertices of $\overleftrightarrow{K}_n$, since the
argument for the other case is similar. For any $e\in
A(\overleftrightarrow{K}_n)\setminus M$, we know $e$ must be
adjacent to at least one element of $M$, so $\lambda_2(D-e)\leq
\min\{\delta^+(D-e), \delta^-(D-e)\}=n-3$ by (\ref{thm2}).
Hence, it suffices to show that $\lambda_2(D)= n-2$ in the
following. We clearly have that $\lambda_2(D)\leq n-2$ by
(\ref{thm2}), so we will show that for $S=\{x, y\}\subseteq V(D)$,
there are at least $n-2$ arc-disjoint strong subgraphs containing
$S$ in $D$.

{\em Case 1.} $x$ and $y$ belong to distinct cycles of
$\overleftrightarrow{K}_n[M]$. We just consider the case that the
lengths of these two cycles are both at least three, since the
arguments for the other cases are similar. Assume that $u_1x, xu_2$
belong to one cycle, and $u_3y, yu_4$ belong to the other cycle.
Note that $u_1u_2, u_3u_4 \in A(D)$ since the lengths of these two
cycles are both at least three.

Let $D_1$ be the 2-cycle $xyx$; let $D_2$ be the subgraph of $D$
with vertex set $\{x, y, u_1, u_2\}$ and arc set $\{xu_1, u_1u_2,
u_2x, yu_2, u_2y\}$; let $D_3$ be the subgraph of $D$ with vertex
set $\{x,y,u_3,u_4\}$ and arc set $\{yu_3, u_3u_4, u_4y, xu_3,
u_3x\}$; let $D_4$ be the subgraph of $D$ with vertex set $\{x, y,
u_1, u_4\}$ and arc set $\{xu_4, u_4x, yu_1, u_1y, u_1u_4,
u_4u_1\}$; for each vertex $u\in V(D)\setminus \{x, y, u_1, u_2,
u_3, u_4\}$, let $D_u$ be a subgraph of $D$ with vertex set $\{u, x
,y\}$ and arc set $\{ux, xu, uy, yu\}$. It is not hard to check that
these $n-2$ strong subgraphs containing $S$ are arc-disjoint.

{\em Case 2.} $x$ and $y$ belong to the same cycle, say $u_1u_2
\cdots u_tu_1$, of $\overleftrightarrow{K}_n[M]$. We just
consider the case that the length of this cycle is at least three,
since the argument for the remaining case is simpler.

{\em Subcase 2.1.} $x$ and $y$ are adjacent in the cycle. Without
loss of generality, let $x=u_1, y=u_2$. Let $D_1$ be the subgraph
of $D$ with vertex set $\{x, y, u_3\}$ and arc set $\{yx, xu_3,
u_3y\}$; let $D_2$ be the subgraph of $D$ with vertex set $\{x, y,
u_3, u_t\}$ and arc set $\{u_3x, xu_t, u_tu_3, u_ty, yu_t\}$; for
each vertex $u\in V(D)\setminus \{x, y, u_3, u_t\}$, let $D_u$ be a
subgraph of $D$ with vertex set $\{u, x ,y\}$ and arc set $\{ux, xu,
uy, yu\}$. It is not hard to check that these $n-2$ strong subgraphs
containing $S$ are arc-disjoint.

{\em Subcase 2.2.} $x$ and $y$ are nonadjacent in the cycle. Without
loss of generality, let $x=u_1, y=u_3$. Let $D_1$ be the 2-cycle
$xyx$; let $D_2$ be the subgraph of $D$ with vertex set $\{x, y,
u_2, u_t\}$ and arc set $\{yu_2, u_2x, xu_t, u_ty\}$; for each
vertex $u\in V(D)\setminus \{x, y, u_2, u_t\}$, let $D_u$ be a
subgraph of $D$ with vertex set $\{u, x ,y\}$ and arc set $\{ux, xu,
uy, yu\}$. It is not hard to check that these $n-2$ strong subgraphs
containing $S$ are arc-disjoint.

\paragraph{Part (b).} Let $H$ be minimally strong subgraph
$(2,n-2)$-arc-connected. By Lemma~\ref{thm2}, we have that $H\not
\cong \overleftrightarrow{K}_n$, that is, $H$ can be obtained from a
complete digraph $\overleftrightarrow{K}_n$ by deleting a nonempty
arc set $M$. To end our argument, we need the following claim. Let
us start from a simple yet useful observation, which follows from (\ref{thm2}).

\begin{pro}\label{pro:HT}
No pair of arcs in $M$ has a common head or tail.
\end{pro}

\vspace{3mm}

Thus, $\overleftrightarrow{K}_n[M]$ must be a union of
vertex-disjoint cycles or paths, otherwise, there are two arcs of
$M$ such that they have a common head or tail, a contradiction with
Proposition \ref{pro:HT}.

\noindent \textbf {Claim 1.} $\overleftrightarrow{K}_n[M]$ does not
contain a path of order at least two.

\noindent {\it Proof of Claim 1.} Let $M'\supseteq M$ be a set of
arcs obtained from $M$ by adding some arcs from
$\overleftrightarrow{K}_n$ such that the digraph
$\overleftrightarrow{K}_n[M']$ contains no path of order at least
two. Note that $\overleftrightarrow{K}_n[M']$ is a supergraph of
$\overleftrightarrow{K}_n[M]$ and is a union of vertex-disjoint
cycles which cover all but at most one vertex of
$\overleftrightarrow{K}_n$. By Part (a), we have that
$\lambda_2(\overleftrightarrow{K}_n[M'])=n-2$, so
$\overleftrightarrow{K}_n[M]$ is not minimally strong subgraph
$(2,n-2)$-arc-connected, a contradiction.

It follows from Claim 1 and its proof that $\overleftrightarrow{K}_n[M]$ must
be a union of vertex-disjoint cycles which cover all but at most one
vertex of $\overleftrightarrow{K}_n$, which completes the proof of Part (b).
\end{pf}

\vskip 1cm 

\noindent {\bf Acknowledgement.}  We are thankful to Anders Yeo for discussions related
 to the complexity of computing strong subgraph $k$-arc-connectivity for semicomplete and symmetric digraphs.

\end{document}